\documentclass[11pt,a4j]{article}

\usepackage[dvipdfmx]{graphicx}

\makeatletter
\def\slash#1{{\mathpalette\c@ncel{#1}}} 
\makeatother

\newcommand\beq{\begin{eqnarray}}
\newcommand\eeq{\end{eqnarray}}

\newcommand\la{\langle}
\newcommand\ra{\rangle}

\begin{document}

\begin{flushright}
\end{flushright}
\vspace*{15mm}
\begin{center}
{\Large \bf New pole contribution to $P_{h\perp}$-weighted single-transverse spin asymmetry 
in semi-inclusive deep inelastic scattering}
\vspace{1.5cm}\\
 {\sc Shinsuke Yoshida$^1$}
\\[0.7cm]
\vspace*{0.1cm}
{\it $^1$ Key Laboratory of Quark and Lepton Physics (MOE) and Institute of Particle Physics,
Central China Normal University, Wuhan 430079, China}\\[3cm]

{\large \bf Abstract} \end{center}

In this paper, we discuss the new hard pole contribution to 
the $P_{h\perp}$-weighted single-transverse spin asymmetry
in semi-inclusive deep inelastic scattering.
We perform the complete next-to-leading order calculation of the $P_{h\perp}$-weighted cross section 
and show that the new hard pole contribution is required in order to 
obtain the complete
evolution equation 
for the Qiu-Sterman function derived 
by 
different approaches.


\newpage
\section{Introduction}

The origin of the single transverse-spin asymmetries(SSAs) in various hard processes 
has been a longstanding problem for almost 40 years since the unexpected 
large asymmetries were observed in mid-1970s~\cite{Klem:1976ui,Bunce:1976yb}.
Many theoretical works in recent decades 
found that twist-3 framework in collinear factorization 
approach is a possible extended framework which can provide a systematic description of 
the large SSA in perturbative QCD. 
The twist-3 framework has been well developed in leading-order(LO) accuracy~[3-17] in recent decades.
Started with the pioneering work by Efremov and Teryaev~[3], 
the more systematic calculation was presented by Qiu and Sterman~[4-6].
While the formalism was applied to SSAs in other processes~\cite{KK2000,JQVY2006,JQVY20062},
the solid foundation was finally provided in~\cite{EKT2007} 
to provide the gauge-invariant twist-3 cross section formula in terms of the 
complete set of the twist-3 distribution functions.
The phenomenological analysis ~\cite{Kouvaris:2006zy,Kanazawa:2010au}
showed that the twist-3 distribution effect of the transversely polarized proton
can give a reasonable description of the experimental data and therefore 
it is widely believed that this effect 
is one of possible sources of the large SSA.


In usual perturbative QCD calculation, higher-order corrections are often not negligible 
compared to a leading order contribution.  
Those corrections
bring the logarithmic energy-scale dependence of nonperturbative function
which is 
described
by the 
evolution equation.
Systematic treatment of the scale dependence of the twist-3 functions
is essential to a quantitative description of the SSA.
The twist-3 distribution effect of the transversely polarized proton is embodied 
as the so-called Qiu-Sterman (QS) function in the spin-dependent cross section formula.
The scale evolution equation of the QS function was 
discussed 
by using several
different approaches so far~[19-26]. One of the approaches is the next-to-leading-order (NLO) 
calculation of the transverse momentum $P_{h\perp}$-weighted cross section. 
Based on this approach, 
some part of the evolution equation was first derived in the study
of the Drell-Yan process~\cite{VY2009}.  
Subsequently
the authors of \cite{KVX2012} examined the
so-called hard-pole (HP) contribution in
the semi-inclusive deep inelastic scattering (SIDIS)
and identified an extra term in the evolution equation which had been derived  
by other approaches~\cite{BMP2009,SZ2012,MW2012,KQ2012},
while the complete agreement for the whole evolution equation was not yet achieved.  
In the meanwhile the authors of \cite{KT2009} found the new HP contribution
in the study of the $P_{h\perp}$-differential cross section for SSA in SIDIS.  
In this paper, we include this new HP contribution for 
the NLO $P_{h}$-weighted cross section. 
We shall show that this new HP contribution yields extra collinear singularity and its
factorization 
reproduce the correct evolution of the QS function found in \cite{BMP2009,MW2012,KQ2012}.
We shall also present the complete NLO cross section for the twist-3 $P_{h}$-weighted cross section for SSA.  

The remainder of the paper is organized as follows: 
in Sec.~2 we introduce the twist-3 distribution functions for the transversely 
polarized proton. Next, in Sec.~3 we discuss the contribution of the real-emission diagrams in NLO
$P_{h\perp}$-weighted cross section. In Sec.~4 we introduce the LO and NLO virtual-correction contributions
which were already calculated in the previous work and present the complete NLO cross section
formula. Finally, in Sec.~5 we summarize our work.


\section{Twist-3 distribution functions for transversely polarized proton}

Here we introduce twist-3 functions relevant to our study.
The F-type twist-3 functions are defined as 
\beq
M^{\alpha}_{F\,ij}(x_1,x_2)&=&\int{d\lambda\over 2\pi}\int{d\mu\over 2\pi}
e^{i\lambda x_1}e^{i\mu(x_2-x_1)}\la pS_{\perp}|\bar{\psi}_j(0)gF^{\alpha n}(\mu n)
\psi_i(\lambda n)|pS_{\perp}\ra \nonumber\\
&=&{M_N\over 4}\epsilon^{\alpha pnS_{\perp}}(\slash{p})_{ij}
G_F(x_1,x_2)+i{M_N\over 4}S^{\alpha}_{\perp}(\gamma_5\slash{p})_{ij}\tilde{G}_F(x_1,x_2)\cdots, 
\label{F-type}
\eeq
where $F^{\alpha n}$ is a gluon's field strength tensor and we used the simplified notation 
$F^{\alpha \beta}n_{\beta}$
and $\epsilon^{\alpha\beta pn} \equiv \epsilon^{\alpha\beta\rho\sigma}p_{\rho}n_{\sigma}$.
The anti-symmetric tensor is defined as $\epsilon^{0123}=-1$.
We introduced the nucleon mass $M_N$ in order to define the dimensionless functions.
From the Hermiticity and $PT$-invariance,
one can show the following symmetry properties:
\beq
G_F(x_1,x_2)=G_F(x_2,x_1),\hspace{5mm}\tilde{G}_F(x_1,x_2)=-\tilde{G}_F(x_2,x_1).
\eeq
In this paper, we discuss the evolution equation of the QS function $G_F(x_1,x_2)$ 
at $x_1=x_2$.


\section{Contribution of real-emission diagrams to next-leading order cross section}
We consider the SSA for light-hadron production 
in SIDIS,
\beq
e(\ell)+p(p,S_{\perp})\to e(\ell')+h(P_h)+X.
\eeq
Within the collinear factorization framework, the SSA can be described
by the twist-3 effects. In this process, the SSA receives two types of twist-3 contributions,
the distribution effect of the transversely polarized proton
and the fragmentation effect of the light-hadron. 
We focus on the former contribution in this study to derive the evolution equation of 
the Qiu-Sterman function $G_F(x,x)$. 
In the case of SIDIS, the cross section formula can be expressed in terms of the following 
Lorentz invariant variables,
\beq
S_{ep}=(p+\ell)^2,\hspace{5mm}Q^2=-q^2,\hspace{5mm}x_B={Q^2\over 2p\cdot q},\hspace{5mm}
z_h={p\cdot P_h\over p\cdot q},
\eeq
where $q=(\ell-\ell')$ is the momentum of the virtual photon. 
We choose the hadron frame~\cite{Kanazawa:2013uia} for the calculation,
\beq
\ell&=&{Q\over 2}(\cosh\psi,\sinh\psi\cos\phi,\sinh\psi\sin\phi,-1),
\\
q&=&(0,0,0,-Q),\hspace{5mm}p^{\mu}=\Bigl({Q\over 2x_B},0,0,{Q\over 2x_B}\Bigr),
\hspace{5mm}S^{\mu}_{\perp}=(0,\cos\Phi_S,\sin\Phi_S,0)
\\
P_h&=&{z_hQ\over 2}\Bigl(1+{P^2_{h\perp}\over z_h^2Q^2}
,{2P_{h\perp}\over z_hQ}\cos\chi,{2P_{h\perp}\over z_hQ}\sin\chi,{P^2_{h\perp}\over z_h^2Q^2}-1\Bigr),
\eeq
where $\cosh\psi={2x_BS_{ep}\over Q^2}-1$.
In this paper, we discuss the NLO
$P_{h\perp}$-weighted polarized cross section defined as
\beq
{d^4\la P_{h\perp}\Delta\sigma\ra\over dx_BdQ^2dz_hd\phi}
\equiv\int d^2P_{h\perp}\epsilon^{S_{\perp}P_{h\perp}pn}
\Bigl({d^6\Delta\sigma\over dx_BdQ^2dz_hdP_{h\perp}^2d\phi d\chi}\Bigr).
\eeq
First we consider the real-emission diagrams in NLO contribution. 
The NLO real-emission diagrams in $P_{h\perp}$-weighted cross section are the same as 
the LO diagrams in $P_{h\perp}$-differential case~\cite{EKT2007,KT2009}. The calculation technique to derive a
twist-3 cross section for $2\to 2$ scattering has been well developed in recent decades
and a systematic way to derive the gauge-invariant cross section 
was established in \cite{EKT2007}. We briefly discuss the derivation below.
The cross section for SIDIS was presented in~\cite{Kanazawa:2013uia,KTY2011} as
\beq
{d^6\Delta\sigma\over dx_BdQ^2dz_hdP_{h\perp}^2d\phi d\chi}
&=&{\alpha^2_{em}\over 128\pi^4z_hx_B^2S^2_{ep}Q^2}L_{\mu\nu}W^{\mu\nu}.
\eeq
where $\alpha_{em}= {e^2\over 4\pi}$ is the QED coupling constant and $L_{\mu\nu}=2(\ell_{\mu}\ell'_{\nu}+\ell_{\nu}\ell'_{\mu})-Q^2g_{\mu\nu}$ is the leptonic 
tensor. Since we are interested in the twist-3 effect of the transversely polarized proton,
we introduce the usual twist-2 fragmentation function $D(z)$ for fragmentation part as
\beq
W^{\mu\nu}&=&\int{{dz\over z^2}}D(z)w^{\mu\nu}
\eeq
The hadronic tensor $w^{\mu\nu}$ describes a scattering of the virtual photon and 
the transversely polarized proton. 
We consider a ``general" diagram given by 
\beq
w^{\mu\nu}&=&\int d^4\xi\int d^4\eta\int{d^4k_1\over (2\pi)^4}
\int{d^4k_2\over (2\pi)^4}e^{ik_1\cdot\xi}e^{i\eta\cdot(k_2-k_1)}
\la PS_{\perp}|\bar{\psi}_j(0)gA_{\alpha}(\eta)\psi_i(\xi)|PS_{\perp}\ra \nonumber\\
&&\times\Bigl(S^{\alpha}_{ji}(k_1,k_2)+\tilde{S}^{\alpha}_{ji}(k_1,k_2)\Bigr),
\label{hadronic tensor}
\eeq
which represents the scattering of the virtual photon and the polarized proton 
graphically shown in Fig.1. We suppressed the Lorentz indices $\mu$ and $\nu$ of 
the hard parts $S^{\alpha}_{ji}(k_1,k_2)$ and $\tilde{S}^{\alpha}_{ji}(k_1,k_2)$ for simplicity.
Within the collinear factorization framework,
a complex phase required for the naively $T$-odd SSA can be provided by a pole contribution 
associated with a internal propagator. In SIDIS case, the pole contributions can be classified into
four types as soft-gluon-pole(SGP), soft-fermion-pole(SFP), hard-pole(HP) 
and another hard-pole(HP2) which are respectively shown in Fig. 2-5. We would like to 
emphasize that the HP2 contribution was not considered in previous studies for 
the $P_{h\perp}$-weighted cross section and this contribution is essential to obtain the consistent evolution 
equation of $G_F(x_1,x_2)$ with the results in different approaches~\cite{BMP2009,MW2012,KQ2012}.
\begin{figure}[h]
\begin{center}
  \includegraphics[height=10cm,width=12cm]{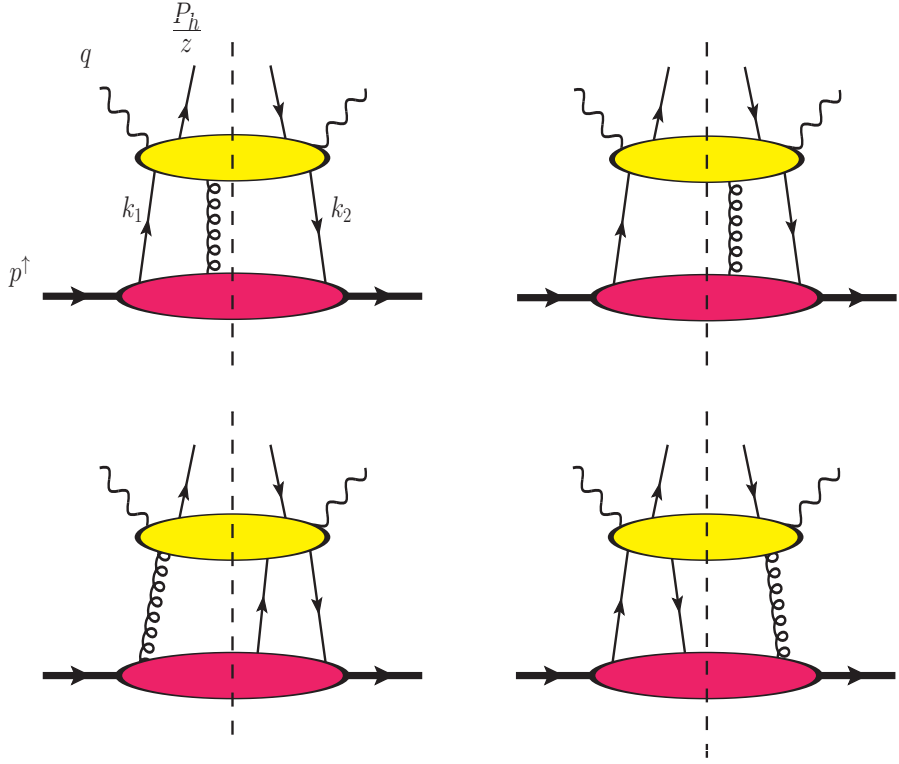}\hspace{1cm}
\end{center}
 \caption{Diagrammatic description for the hadronic tensor $w^{\mu\nu}$. 
 The upper diagrams and the lower diagrams respectively
 represent $S^{\alpha}_{ji}(k_1,k_2)$ and $\tilde{S}^{\alpha}_{ji}(k_1,k_2)$.
 }
\end{figure}
\begin{figure}[h]
\begin{center}
  \includegraphics[height=8cm,width=12cm]{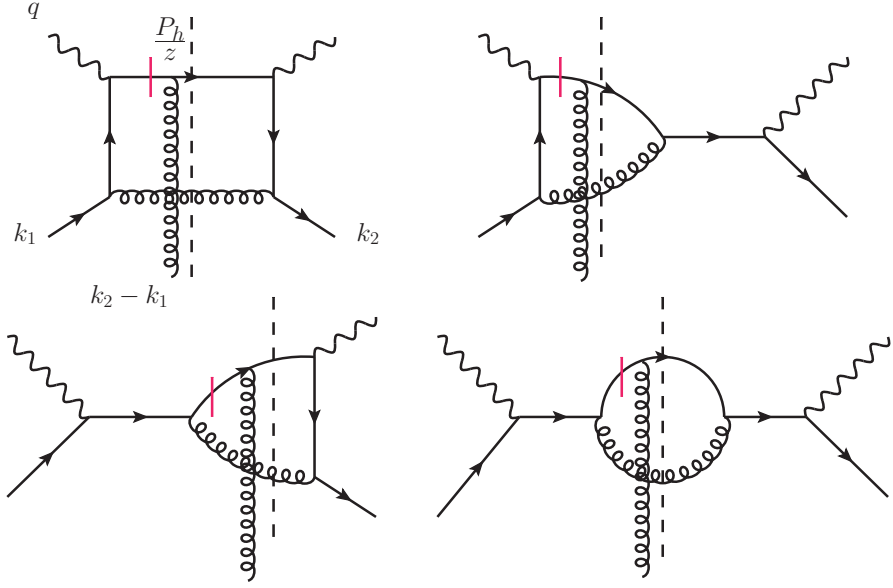}\hspace{1cm}
\end{center}
 \caption{Diagrammatic description for SGP diagrams $H^{\rm SGP\,\alpha}_{Lji}(k_1,k_2)$. 
 Barred propagators provide the pole contribution. 
 }
\end{figure}
\begin{figure}[h]
\begin{center}
  \includegraphics[height=4cm,width=12cm]{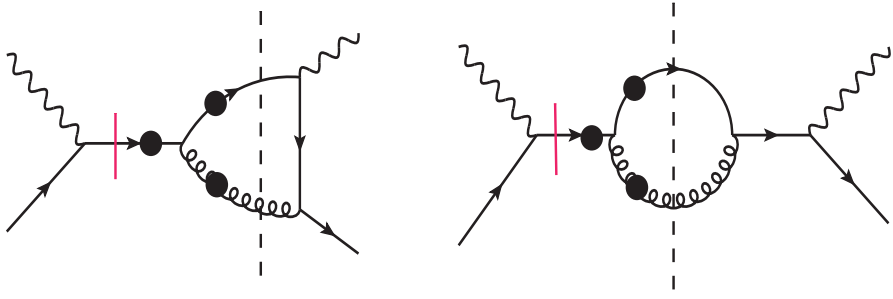}\hspace{1cm}
\end{center}
 \caption{Diagrammatic description for HP diagrams $H^{\rm HP\,\alpha}_{Lji}(k_1,k_2)$.
 The third gluon line with momentum $k_2-k_1$ which comes from the transversely polarized proton
 attaches to one of the black dots in each diagram.}
\end{figure}
\begin{figure}[h]
\begin{center}
  \includegraphics[height=8cm,width=12cm]{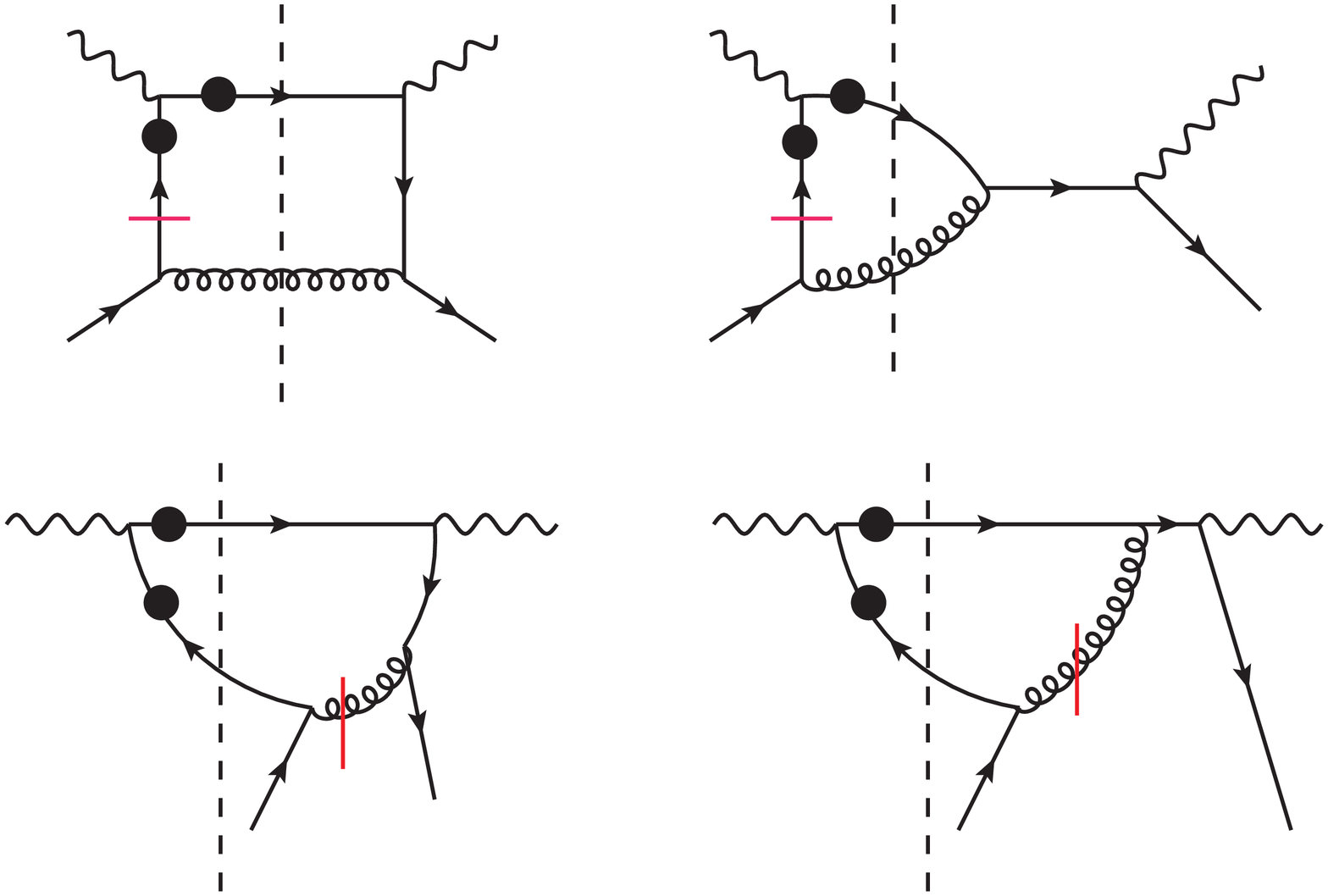}\hspace{1cm}
\end{center}
 \caption{Diagrammatic description for SFP diagrams. 
 The upper diagrams and the lower diagrams respectively
 represent $H^{{\rm SFP}\,\alpha}_{ji}(k_1,k_2)$ and $\tilde{H}^{{\rm SFP}\,\alpha}_{ji}(k_1,k_2)$.
 }
\end{figure}
\begin{figure}[h]
\begin{center}
  \includegraphics[height=3cm,width=12cm]{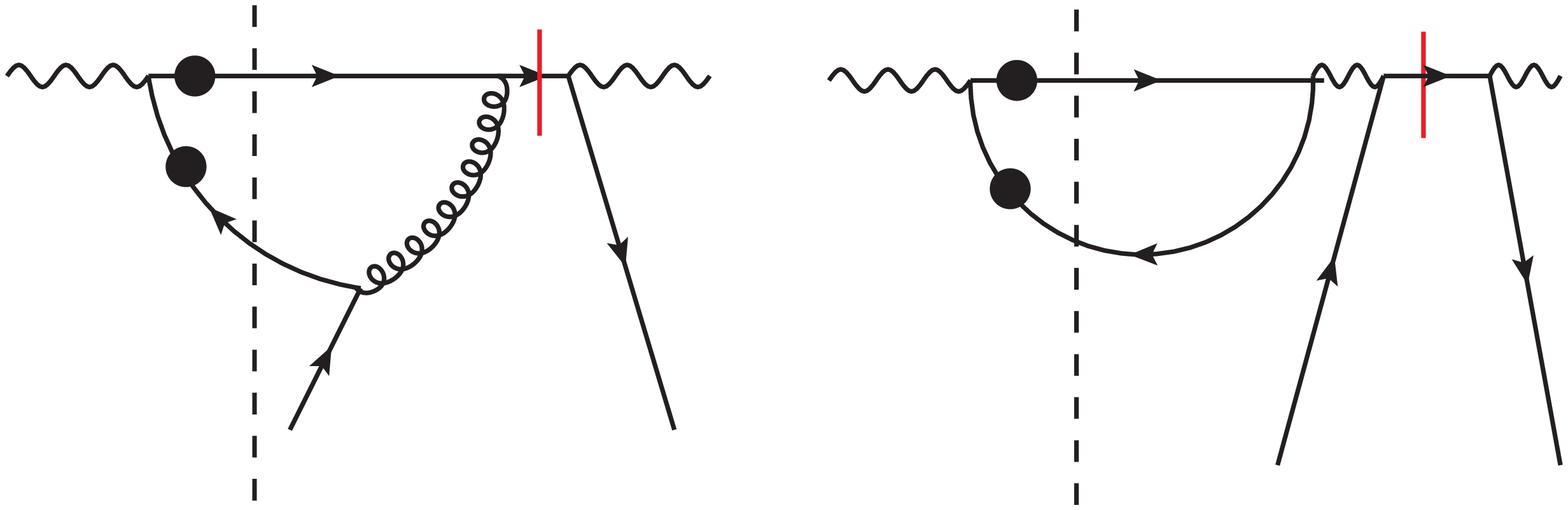}\hspace{1cm}
\end{center}
 \caption{Diagrammatic description for HP2 diagrams $H^{\rm HP2\,\alpha}_{Lji}(k_1,k_2)$. 
These diagrams were first found in \cite{KT2009} in the study of $P_{h\perp}$-differential
SSA but was not considered in previous studies of the $P_{h\perp}$-weighted SSA.}
\end{figure}
We can check that the hard part $S^{\alpha}_{ji}(k_1,k_2)$ with the pole contribution 
satisfies the Ward identity
\beq
(k_2-k_1)_{\alpha}S^{{\rm pole}\ \alpha}_{ji}(k_1,k_2)&=&0,
\label{ward}
\eeq
and associated relations 
\beq
(x_2-x_1){\partial\over \partial k_2^{\alpha}}S^{{\rm pole}\ p}_{ji}(k_1,k_2)\Bigr|_{k_i=x_ip}&=&
-S^{{\rm pole}\ \alpha}_{ji}(x_1p,x_2p),
\label{ward2}\\
(x_2-x_1){\partial\over \partial k_1^{\alpha}}S^{{\rm pole}\ p}_{ji}(k_1,k_2)\Bigr|_{k_i=x_ip}&=&
S^{{\rm pole}\ \alpha}_{ji}(x_1p,x_2p).
\eeq
For SFP and HP contributions, the above relations give
\beq
{\partial\over \partial k_2^{\alpha}}S^{{\rm pole}\ p}_{ji}(k_1,k_2)\Bigr|_{k_i=x_ip}
=-{\partial\over \partial k_1^{\alpha}}S^{{\rm pole}\ p}_{ji}(k_1,k_2)\Bigr|_{k_i=x_ip},
\label{ward3}
\eeq
and we can find the same relation for SGP contribution with a direct inspection.
Another hard part $\tilde{S}^{{\rm pole}\ p}_{ji}(k_1,k_2)$ also has the same relations.
To extract the twist-3 $O(k_{\perp})$ contribution from the general contribution (\ref{hadronic tensor}),
we perform the collinear expansion for the hard parts as
\beq
S^{{\rm pole}\ \alpha}_{ji}(k_1,k_2)&=&
S^{{\rm pole}\ \alpha}_{ji}((k_1\cdot n)p,(k_2\cdot n)p)
+{\partial\over \partial k_1^{\alpha}}S^{{\rm pole}\ p}_{ji}
(k_1,k_2)\Bigr|_{k_i=(k_i\cdot n)p}\omega^{\alpha}_{\ \beta}k_1^{\beta} \nonumber\\
&&+{\partial\over \partial k_2^{\alpha}}S^{{\rm pole}\ p}_{ji}
(k_1,k_2)\Bigr|_{k_i=(k_i\cdot n)p}\omega^{\alpha}_{\ \beta}k_2^{\beta} \nonumber\\
&=&S^{{\rm pole}\ \alpha}_{ji}((k_1\cdot n)p,(k_2\cdot n)p)
+{\partial\over \partial k_2^{\alpha}}S^{{\rm pole}\ p}_{ji}
(k_1,k_2)\Bigr|_{k_i=(k_i\cdot n)p}\omega^{\alpha}_{\ \beta}(k_2^{\beta}-k_1^{\beta}),
\hspace{5mm}
\eeq
where $\omega^{\alpha}_{\ \beta}=g^{\alpha}_{\ \beta}-p^{\alpha}n_{\beta}$
and we used the relation (\ref{ward3}). And we separate the Lorentz components of the gluon 
field,
\beq
A^{\alpha}=A^{n}p^{\alpha}+\omega^{\alpha}_{\ \beta}A^{\beta}.
\eeq
Then we pick up subleading contributions in (\ref{hadronic tensor}) and
construct the F-type correlator (\ref{F-type}) as
\beq
w^{\mu\nu}&=&\int d^4\xi\int d^4\eta\int{d^4k_1\over (2\pi)^4}
\int{d^4k_2\over (2\pi)^4}e^{ik_1\cdot\xi}e^{i\eta\cdot(k_2-k_1)}
\la PS_{\perp}|\bar{\psi}_j(0)gA^n(\eta)\psi_i(\xi)|PS_{\perp}\ra 
\nonumber\\
&&\times 
{\partial\over \partial k_2^{\alpha}}\Bigl(S^{{\rm pole}\ p}_{ji}
(k_1,k_2)+\tilde{S}^{{\rm pole}\ p}_{ji}(k_1,k_2)\Bigr)
\Bigr|_{k_i=(k_i\cdot n)p}\omega^{\alpha}_{\ \beta}(k_2^{\beta}-k_1^{\beta})
\nonumber\\
&&+\int d^4\xi\int d^4\eta\int{d^4k_1\over (2\pi)^4}
\int{d^4k_2\over (2\pi)^4}e^{ik_1\cdot\xi}e^{i\eta\cdot(k_2-k_1)}
\la PS_{\perp}|\bar{\psi}_j(0)g\omega_{\alpha}^{\ \beta}A_{\beta}(\eta)\psi_i(\xi)|PS_{\perp}\ra \nonumber\\
&&\times\Bigl(S^{{\rm pole}\ p}_{ji}((k_1\cdot n)p,(k_2\cdot n)p)
+\tilde{S}^{{\rm pole}\ p}_{ji}((k_1\cdot n)p,(k_2\cdot n)p)\Bigr)
\nonumber\\
&=&i\omega_{\alpha}^{\ \beta}\int dx_1\int dx_2 M^{\alpha}_{ij\,F}(x_1,x_2)
{\partial\over \partial k_2^{\beta}}\Bigl(S^{{\rm pole}\ p}_{ji}
(k_1,k_2)+\tilde{S}^{{\rm pole}\ p}_{ji}(k_1,k_2)\Bigr)
\Bigr|_{k_i=x_ip}
\eeq
We express the hard parts in terms of each pole contribution as
\beq
S^{{\rm pole}\,\alpha}_{ji}(k_1,k_2)&=&
H^{{\rm SGP}\,\alpha}_{L\,ji}(k_1,k_2)\Bigl\{-i\pi\delta\Bigl(({P_h\over z}-(k_2-k_1))^2\Bigr)\Bigr\}
(2\pi)\delta\Bigl((k_2+q-{P_h\over z})^2\Bigr)
\nonumber\\
&&+H^{{\rm HP}\,\alpha}_{L\,ji}(k_1,k_2)\Bigl\{-i\pi\delta\Bigl((k_1+q)^2\Bigr)\Bigr\}
(2\pi)\delta\Bigl((k_2+q-{P_h\over z})^2\Bigr)
\nonumber\\
&&+H^{{\rm SFP}\,\alpha}_{L\,ji}(k_1,k_2)\Bigl\{-i\pi\delta\Bigl(({P_h\over z}-(k_2-k_1)-q)^2\Bigr)\Bigr\}
(2\pi)\delta\Bigl((k_2+q-{P_h\over z})^2\Bigr)
\nonumber\\
&&+{\rm mirror\ diagrams}
\\
\tilde{S}^{{\rm pole}\,\alpha}_{ji}(k_1,k_2)&=&
\tilde{H}^{{\rm HP2}\,\alpha}_{L\,ji}(k_1,k_2)\Bigl\{i\pi\delta\Bigl((k_2+q)^2\Bigr)\Bigr\}
(2\pi)\delta\Bigl((k_2-k_1+q-{P_h\over z})^2\Bigr)
\nonumber\\
&&+\tilde{H}^{{\rm SFP}\,\alpha}_{L\,ji}(k_1,k_2)\Bigl\{i\pi\delta\Bigl(({P_h\over z}-k_2-q)^2\Bigr)\Bigr\}
(2\pi)\delta\Bigl((k_2-k_1+q-{P_h\over z})^2\Bigr)
\nonumber\\
&&+{\rm mirror\ diagrams}
\eeq
We can find that the SFP contributions $H^{{\rm SFP}\ p}_{ji}(k_1,k_2)$ and 
$\tilde{H}^{{\rm SFP}\ p}_{ji}(k_1,k_2)$ are the topologically same and then exactly 
cancel each other.
After a little computation, we can obtain the formula for the hadronic tensor $W^{\mu\nu}$ as follows.
\beq
W^{\mu\nu}
&=&{M_N\pi^2\over 2}\int{dz\over z^2}D(z)\int {dx\over x}\delta\Bigl((xp+q-{P_h\over z})^2\Bigr)
\Bigl[-2\epsilon^{p_cpnS_{\perp}}{d\over dx}G_F(x,x){\hat{s}+Q^2\over \hat{t}\hat{u}}{\rm Tr}[x\slash{p}H(xp)]
\nonumber\\
&&-2\epsilon^{p_cpnS_{\perp}}G_F(x,x){\hat{s}+Q^2\over \hat{t}\hat{u}}\Bigl\{
Q^2\Bigl({\partial\over \partial \hat{s}}-{\partial\over \partial Q^2}\Bigr){\rm Tr}[x\slash{p}H(xp)]
\Bigr\}
\nonumber\\
&&+G_F(x,x_B){1\over \hat{x}-1}{\hat{x}\over Q^2}\epsilon_{\alpha}^{\ pnS_{\perp}}
\Bigl({\rm Tr}[x\slash{p}H_L^{{\rm HP}\,\alpha}(x_Bp,xp)]
+{\rm Tr}[x\slash{p}H_R^{{\rm HP}\,\alpha}(xp,x_Bp)]
\Bigr)
\nonumber\\
&&-\tilde{G}_F(x,x_B){1\over \hat{x}-1}{\hat{x}\over Q^2}iS_{\perp\alpha}
\Bigl({\rm Tr}[\gamma_5x\slash{p}\tilde{H}_L^{{\rm HP}\,\alpha}(x_Bp,xp)]
-{\rm Tr}[\gamma_5x\slash{p}\tilde{H}_R^{{\rm HP}\,\alpha}(xp,x_Bp)]
\nonumber\\
&&+G_F(x_B,x_B-x){\hat{x}\over Q^2}\epsilon_{\alpha}^{\ pnS_{\perp}}
\Bigl({\rm Tr}[x\slash{p}H_L^{{\rm HP2}\,\alpha}((x_B-x)p,x_Bp)]
\nonumber\\
&&
+{\rm Tr}[x\slash{p}H_R^{{\rm HP2}\,\alpha}(x_Bp,(x_B-x)p)]
\Bigr)
\nonumber\\
&&-\tilde{G}_F(x_B,x_B-x){\hat{x}\over Q^2}iS_{\perp\alpha}
\Bigl({\rm Tr}[\gamma_5x\slash{p}\tilde{H}_L^{{\rm HP2}\,\alpha}((x_B-x)p,x_Bp)]
\nonumber\\
&&-{\rm Tr}[\gamma_5x\slash{p}\tilde{H}_R^{{\rm HP2}\,\alpha}(x_Bp,(x_B-x)p)]
\Bigr],
\eeq
where we used the Mandelstam variables
\beq
\hat{s}&=&(xp+q)^2={1-\hat{x}\over \hat{x}}Q^2,
\\
\hat{t}&=&(p_c-q)^2=-{1-\hat{z}\over \hat{x}}Q^2,
\\
\hat{u}&=&(xp-p_c)^2=-{\hat{z}\over \hat{x}}Q^2,
\eeq
where $p_c={P_h\over z}$.
We used the Ward identity (\ref{ward2}) for hard-pole contributions.
For the SGP contribution, we used master formula~\cite{KTY2011,KT2006}
\beq
{\partial\over \partial k_2^{\beta}}{\rm Tr}[x_1\slash{p}
S^{{\rm SGP}\ p}_{ji}(k_1,k_2)]\Bigr|_{k_i=x_ip}
&=&-i\pi\delta(x_1-x_2){d\over dp^{\beta}_c}{\rm Tr}[x_1\slash{p}S(x_1p)]
\nonumber\\
&=&2i\pi\delta(x_1-x_2){\hat{s}+Q^2\over \hat{u}}p^{\beta}_c
{\partial\over \partial \hat{t}}{\rm Tr}[x_1\slash{p}S(x_1p)],
\eeq
where $S(xp)$ is the $2\to 2$ scattering cross section without the third gluon line comes from 
the transversely polarized proton
(but the color factor is the same as $S^{SGP\,p}_{ji}$).
In this paper, we consider the metric contribution,
\beq
L_{\mu\nu}W^{\mu\nu}\to (-g_{\mu\nu}W^{\mu\nu}),
\eeq
\beq
{d^4\la P_{h\perp}\Delta\sigma\ra^{\rm real}\over dx_BdQ^2dz_hd\phi}
={\alpha^2_{em}\over 32\pi^2z_hx_B^2S^2_{ep}Q^2}\int{dz}zD(z)
\int {d^{2}p_{c\perp}\over (2\pi)^{2}}\epsilon^{S_{\perp}p_{c\perp}pn}
\Bigl(-g_{\mu\nu}w^{\mu\nu}\Bigr),
\eeq
and the metric should be normalized as $\displaystyle g_{\mu\nu}\to {1\over 1-\epsilon}g_{\mu\nu}$
with $\epsilon=2-D/2$ in $D$-dimensional calculation.
We can compute the $P_{h\perp}$-weighted cross section for NLO real-emission diagrams 
in $D$-dimension as follows. 
\beq
&&{d^4\la P_{h\perp}\Delta\sigma\ra^{\rm real}\over dx_BdQ^2dz_hd\phi}
\nonumber\\
&=&-{\pi M_N\alpha^2_{em}\alpha_s\over 2 x_B^2S^2_{ep}Q^2}\sum_qe^2_q\int{dz}D^q(z)
\mu^{2\epsilon}\int {d^{2-2\epsilon}p_{c\perp}\over (2\pi)^{2-2\epsilon}}
\Bigl[\int {dx\over x}
\delta\Bigl(p^2_{c\perp}-{(1-\hat{x})(1-\hat{z})\hat{z}\over \hat{x}}Q^2\Bigr) 
\nonumber\\
&&\times{1\over 1-\epsilon}\Bigl[{d\over dx}G^q_F(x,x)H_D 
+G^q_F(x,x)H_{ND}+G^q_F(x,x_B)H_{HP}
+\tilde{G}^q_F(x,x_B)H_{HPT}
\nonumber\\
&&+G^q_F(x_B,x_B-x)H_{HP2}
+\tilde{G}^q_F(x_B,x_B-x)H_{HPT2}
\Bigr],
\eeq
where $q$ denotes the quark flavor, 
$\alpha_s$ is the QCD coupling constant and we used the symmetry for $p_{c\perp}$-integral
\beq
\int d^{2-2\epsilon}p_{c\perp}\,p_{c\perp\alpha}p_{c\perp\beta}\epsilon^{S_{\perp}\alpha pn}
\epsilon^{\beta pnS_{\perp}}&=&
-\int d^{2-2\epsilon}p_{c\perp}\,{1\over 2(1-\epsilon)}p^2_{c\perp}g_{\perp\alpha\beta}
\epsilon^{S_{\perp}\alpha pn}\epsilon^{\beta pnS_{\perp}}
\nonumber\\
&=&-\int d^{2-2\epsilon}p_{c\perp}\,{1\over 2(1-\epsilon)}{(1-\hat{x})(1-\hat{z})\hat{z}\over \hat{x}}Q^2,
\eeq
and the hard cross sections can be computed as
\beq
H_D&=&{1\over 2N}\Bigl\{1-2\hat{x}-\hat{z}
+\epsilon(1-2\hat{x}+\hat{z})
+{1+\hat{x}^2
-\epsilon(1-\hat{x})^2\over 1-\hat{z}}
\Bigr\}
\\
H_{ND}&=&{1\over 2N}\Bigl[-{2\over (1-\hat{x})(1-\hat{z})}+{1+\hat{z}+\epsilon(1-\hat{z})\over 1-\hat{x}}
\nonumber\\
&&+{(1-\hat{x})(1+2\hat{x})-\epsilon(1-\hat{x})(2\hat{x}-1)\over 1-\hat{z}}
-2(1+\epsilon)(1-\hat{x})\Bigr]
\\
H_{HP}&=&\Bigl(\hat{z}C_F+{1\over 2N}\Bigr)\Bigl[{2\over(1-\hat{x})(1-\hat{z})}-{1+\hat{z}+\epsilon(1-\hat{z})\over 1-\hat{x}}
-{1\over 1-\hat{z}}+(1+\hat{z}+\epsilon)\Bigr]
\\
H_{HPT}&=&{1\over 1-\epsilon}\Bigl(\hat{z}C_F+{1\over 2N}\Bigr)\Bigl[
-{1+\hat{z}-2\epsilon+\epsilon^2(1-\hat{z})\over 1-\hat{x}}
-{1+\epsilon\over 1-\hat{z}}+(1+\hat{z}+\epsilon\hat{z}+\epsilon^2)\Bigr]
\\
H_{HP2}&=&{1\over 2N}\Bigl[{1-2\hat{x}\over 1-\hat{z}}-(1-2\hat{x})(1+\hat{z}+\epsilon)\Bigr]
+{1\over 1-\epsilon}{1\over 2}\Bigl[(1-2\hat{x})(2\hat{z}^2-2\hat{z}+1-\epsilon)\Bigr]
\\
H_{HPT2}&=&{1\over 1-\epsilon}{1\over 2N}\Bigl[{1+\epsilon\over 1-\hat{z}}
-(1+\hat{z}+\epsilon\hat{z}+\epsilon^2)\Bigr]
-{1\over 1-\epsilon}{1\over 2}\Bigl[1-2\hat{z}-\epsilon\Bigr],
\eeq
where $N=3$ is a number of colors and $C_F={N^2-1\over 2N}$.
The $p_{c\perp}$-integral can be calculated in $D$-dimension as 
\beq
&&\int{d^{2-2\epsilon}p_c\over (2\pi)^{2-2\epsilon}}\delta
\Bigl(p^2_{c\perp}-{(1-\hat{x})(1-\hat{z})\hat{z}\over \hat{x}}Q^2\Bigr)
\nonumber\\
&=&{1\over (2\pi)^{2-2\epsilon}}\int{dp_{c\perp}}\int d\Omega_{2-2\epsilon}
(p_{c\perp})^{1-2\epsilon}\delta
\Bigl(p^2_{c\perp}-{(1-\hat{x})(1-\hat{z})\hat{z}\over \hat{x}}Q^2\Bigr)
\nonumber\\
&=&{1\over 4\pi}\Bigl({4\pi\over Q^2}\Bigr)^{\epsilon}{1\over \Gamma(1-\epsilon)}
\Bigl({(1-\hat{x})(1-\hat{z})\hat{z}\over \hat{x}}\Bigr)^{-\epsilon},
\eeq
where $\Omega_{2-2\epsilon}$ is a solid angle
\beq
\int d\Omega_{2-2\epsilon}={2\pi^{1-\epsilon}\over \Gamma(1-\epsilon)}.
\eeq
We carry out the $\epsilon$-expansion for the phase-space integral as follows.
\beq
\hat{z}^{-\epsilon}&\simeq& 1-\epsilon\ln\hat{z},\hspace{10mm}
\hat{x}^{\epsilon}\simeq 1+\epsilon\ln\hat{z},
\\
(1-\hat{z})^{-1-\epsilon}&\simeq& -{1\over \epsilon}\delta(1-\hat{z})+{1\over (1-\hat{z})_+}
-\epsilon\Bigl({\ln(1-\hat{z})\over 1-\hat{z}}\Bigr)_+,
\\
(1-\hat{x})^{-1-\epsilon}&\simeq& -{1\over \epsilon}\delta(1-\hat{x})+{1\over (1-\hat{x})_+}
-\epsilon\Bigl({\ln(1-\hat{x})\over 1-\hat{x}}\Bigr)_+,
\eeq
Then the cross section formula reads
\beq
&&{d^4\la P_{h\perp}\Delta\sigma\ra^{\rm real}\over dx_BdQ^2dz_hd\phi}
\nonumber\\
&=&-{\pi M_N\alpha^2_{em}\over 4x_B^2S^2_{ep}Q^2}{\alpha_s\over 2\pi}
\Bigl({4\pi\mu^2\over Q^2}\Bigr)^{\epsilon}{1\over \Gamma(1-\epsilon)}
\sum_qe_q^2\Bigl[\int dzD^q(z)\int {dx\over x}
\Bigl[{d\over dx}G^q_F(x,x)\hat{\sigma}_D 
+G^q_F(x,x)\hat{\sigma}_{ND}
\nonumber\\
&&
+G^q_F(x,x_B)\hat{\sigma}_{HP}
+\tilde{G}^q_F(x,x_B)\hat{\sigma}_{HPT}
+G^q_F(x_B,x_B-x)\hat{\sigma}_{HP2}
+\tilde{G}^q_F(x_B,x_B-x)\hat{\sigma}_{HPT2}
\Bigr],
\nonumber\\
\eeq
\beq
\hat{\sigma}_D&=&{1\over 2N}\Bigl[(-{1\over \epsilon})(1+\hat{x}^2)\delta(1-\hat{z})
+(1-\hat{z})+{(1-\hat{x})^2+2\hat{x}\hat{z}\over (1-\hat{z})_+}
\nonumber\\
&&-\delta(1-\hat{z})\Bigl((1+\hat{x}^2)\ln{\hat{x}\over 1-\hat{x}}+2\hat{x}\Bigr)
\Bigr]
\label{hard1}\\
\hat{\sigma}_{ND}&=&{1\over 2N}\Bigl[(-{2\over \epsilon^2})\delta(1-\hat{x})\delta(1-\hat{z})
+(-{1\over \epsilon})\Bigl(2\delta(1-\hat{x})\delta(1-\hat{z})
-{1+\hat{z}^2\over (1-\hat{z})_+}\delta(1-\hat{x})
\nonumber\\
&&
+{2\hat{x}^3-3\hat{x}^2-1\over (1-\hat{x})_+}\delta(1-\hat{z})\Bigr)
-2\delta(1-\hat{x})\delta(1-\hat{z})+{2\hat{x}^3-3\hat{x}^2-1\over (1-\hat{x})_+(1-\hat{z})_+}
\nonumber\\
&&+{1+\hat{z}\over (1-\hat{x})_+}-2(1-\hat{x})
+\delta(1-\hat{z})\Bigl(
-(1-\hat{x})(1+2\hat{x})\log{\hat{x}\over 1-\hat{x}}-2\Bigl({\ln(1-\hat{x})\over 1-\hat{x}}\Bigr)_+
\nonumber\\
&&+{2\over (1-\hat{x})_+}
-2(1-\hat{x})
+2{\ln\hat{x}\over (1-\hat{x})_+}
\Bigr)
+\delta(1-\hat{x})\Bigl(
(1+\hat{z})\ln\hat{z}(1-\hat{z})-2{\ln\hat{z}\over (1-\hat{z})_+}
\nonumber\\
&&-2\Bigl({\ln(1-\hat{z})\over 1-\hat{z}}\Bigr)_+
+{2\hat{z}\over (1-\hat{z})_+}\Bigr)\Bigr]
\\
\hat{\sigma}_{HP}&=&\Bigl(\hat{z}C_F+{1\over 2N}\Bigr)\Bigl[{2\over \epsilon^2}\delta(1-\hat{x})\delta(1-\hat{z})
+{1\over \epsilon}\Bigl(2\delta(1-\hat{x})\delta(1-\hat{z})
-{1+\hat{z}^2\over (1-\hat{z})_+}\delta(1-\hat{x})
\nonumber\\
&&-{1+\hat{x}\over (1-\hat{x})_+}\delta(1-\hat{z})
\Bigr)
+2\delta(1-\hat{x})\delta(1-\hat{z})+{1+\hat{x}\hat{z}^2\over (1-\hat{x})_+(1-\hat{z})_+}
\nonumber\\
&&+\delta(1-\hat{z})\Bigl(
\log{\hat{x}\over 1-\hat{x}}+2\Bigl({\ln(1-\hat{x})\over 1-\hat{x}}\Bigr)_+
-2{\ln\hat{x}\over (1-\hat{x})_+}-{1+\hat{x}\over (1-\hat{x})_+}
\Bigr)
\nonumber\\
&&+\delta(1-\hat{x})\Bigl(
-(1+\hat{z})\ln\hat{z}(1-\hat{z})+2\Bigl({\ln(1-\hat{z})\over 1-\hat{z}}\Bigr)_+
+2{\ln\hat{z}\over (1-\hat{z})_+}-{2\hat{z}\over (1-\hat{z})_+}
\Bigr)
\Bigr]
\label{hard2}
\\
\hat{\sigma}_{HPT}&=&\Bigl(\hat{z}C_F+{1\over 2N}\Bigr)\Bigl[{1\over \epsilon}\delta(1-\hat{z})
-{1-\hat{x}\hat{z}^2\over (1-\hat{x})_+(1-\hat{z})_+}
+\delta(1-\hat{z})\Bigl(
\ln{\hat{x}\over 1-\hat{x}}+3
\Bigr)\Bigr]
\label{hard3}
\\
\hat{\sigma}_{HP2}&=&{1\over 2N}\Bigl[
-{1\over \epsilon}(1-2\hat{x})\delta(1-\hat{z})
+{(1-2\hat{x})\hat{z}^2\over (1-\hat{z})_+}
-\delta(1-\hat{z})(1-2\hat{x})
(\ln{\hat{x}\over 1-\hat{x}}+1)
\Bigr]
\nonumber\\
&&+{1\over 2}(1-2\hat{x})((1-\hat{z})^2+\hat{z}^2)
\\
\hat{\sigma}_{HPT2}&=&{1\over 2N}\Bigl[
-{1\over \epsilon}\delta(1-\hat{z})
+{\hat{z}^2\over (1-\hat{z})_+}
-\delta(1-\hat{z})(\ln{\hat{x}\over 1-\hat{x}}+3)
\Bigr]
-{1\over 2}(1-2\hat{z}),\label{hard4}
\eeq
where we used the antisymmetric property $\tilde{G}_F(x,x_B)\delta(1-\hat{x})=\tilde{G}_F(x,x)\delta(1-\hat{x})=0$.
Finally we can derive the contribution of real-emission diagrams as
\beq
&&{d^4\la P_{h\perp}\Delta\sigma\ra^{\rm real}\over dx_BdQ^2dz_hd\phi}
\nonumber\\
&=&-{z_h\pi M_N\alpha^2_{em}\over 4x_B^2S^2_{ep}Q^2}{\alpha_s\over 2\pi}
\Bigl({4\pi\mu^2\over Q^2}\Bigr)^{\epsilon}{1\over \Gamma(1-\epsilon)}
\sum_qe_q^2\Biggl[C_F{2\over \epsilon^2}G^q_F(x_B,x_B)D^q(z_h)
\nonumber\\
&&+\Bigl(-{1\over \epsilon}\Bigr)\Biggl\{
D^q(z_h)\Bigl\{\int^1_{x_B}{dx\over x}\Bigl[C_F{1+\hat{x}^2\over (1-\hat{x})_+}G^q_F(x,x)
+{N\over 2}\Bigl({(1+\hat{x})G^q_F(x_B,x)-(1+\hat{x}^2)G^q_F(x,x)\over (1-\hat{x})_+}
\nonumber\\
&&+\tilde{G}^q_F(x_B,x)\Bigr)\Bigr]
-NG^q_F(x_B,x_B)
+{1\over 2N}\int^1_{x_B}{dx\over x}\Bigl((1-2\hat{x})G^q_F(x_B,x_B-x)+\tilde{G}^q_F(x_B,x_B-x)\Bigr)
\Bigr\}
\nonumber\\
&&+G^q_F(x_B,x_B)C_F
\int^1_{z_h}{dz\over z}{1+\hat{z}^2\over (1-\hat{z})_+}D^q(z)\Biggr\}
\nonumber\\
&&+\int^1_{x_B}{dx\over x}\int^1_{z_h}{dz\over z}
\Biggl\{x{dx\over x}G^q_F(x,x)D^q(z)
{1\over 2N\hat{z}}\Bigl[1-\hat{z}+{(1-\hat{x})^2+2\hat{x}\hat{z}\over (1-\hat{z})_+}
\nonumber\\
&&-\delta(1-\hat{z})\Bigl((1+\hat{x}^2)\ln{\hat{x}\over 1-\hat{x}}+2\hat{x}\Bigr)\Bigr]
+G^q_F(x,x)D^q(z){1\over 2N\hat{z}}\Bigl[
-2\delta(1-\hat{x})\delta(1-\hat{z})
\nonumber\\
&&+{2\hat{x}^3-3\hat{x}^2-1\over (1-\hat{x})_+(1-\hat{z})_+}
+{1+\hat{z}\over (1-\hat{x})_+}-2(1-\hat{x})
+\delta(1-\hat{z})\Bigl(
-(1-\hat{x})(1+2\hat{x})\log{\hat{x}\over 1-\hat{x}}
\nonumber\\
&&-2\Bigl({\ln(1-\hat{x})\over 1-\hat{x}}\Bigr)_+
+{2\over (1-\hat{x})_+}
-2(1-\hat{x})
+2{\ln\hat{x}\over (1-\hat{x})_+}
\Bigr)
+\delta(1-\hat{x})\Bigl(
(1+\hat{z})\ln\hat{z}(1-\hat{z})
\nonumber\\
&&-2{\ln\hat{z}\over (1-\hat{z})_+}
-2\Bigl({\ln(1-\hat{z})\over 1-\hat{z}}\Bigr)_+
+{2\hat{z}\over (1-\hat{z})_+}\Bigr)
\Bigr]
+G^q_F(x,x_B)D^q(z)\Bigl(C_F+{1\over 2N\hat{z}}\Bigr)\Bigl[
2\delta(1-\hat{x})\delta(1-\hat{z})
\nonumber\\
&&+{1+\hat{x}\hat{z}^2\over (1-\hat{x})_+(1-\hat{z})_+}
+\delta(1-\hat{z})\Bigl(
\log{\hat{x}\over 1-\hat{x}}+2\Bigl({\ln(1-\hat{x})\over 1-\hat{x}}\Bigr)_+
-2{\ln\hat{x}\over (1-\hat{x})_+}-{1+\hat{x}\over (1-\hat{x})_+}
\Bigr)
\nonumber\\
&&+\delta(1-\hat{x})\Bigl(
-(1+\hat{z})\ln\hat{z}(1-\hat{z})+2\Bigl({\ln(1-\hat{z})\over 1-\hat{z}}\Bigr)_+
+2{\ln\hat{z}\over (1-\hat{z})_+}-{2\hat{z}\over (1-\hat{z})_+}
\Bigr)
\Bigr]
\nonumber\\
&&+\tilde{G}^q_F(x,x_B)D^q(z)\Bigl(C_F+{1\over 2N\hat{z}}\Bigr)\Bigl[
-{1-\hat{x}\hat{z}^2\over (1-\hat{x})_+(1-\hat{z})_+}
+\delta(1-\hat{z})\Bigl(
\ln{\hat{x}\over 1-\hat{x}}+3
\Bigr)
\Bigr]
\nonumber\\
&&+G^q_F(x_B,x_B-x)D^q(z)\Bigl[{1\over 2N\hat{z}}\Bigl(
{(1-2\hat{x})\hat{z}^2\over (1-\hat{z})_+}
-\delta(1-\hat{z})(1-2\hat{x})
(\ln{\hat{x}\over 1-\hat{x}}+1)\Bigr)
\nonumber\\
&&+{1\over 2\hat{z}}(1-2\hat{x})\{(1-\hat{z})^2+\hat{z}^2\}
\Bigr]+\tilde{G}^q_F(x_B,x_B-x)D^q(z)\Bigl[{1\over 2N\hat{z}}\Bigl(
{\hat{z}^2\over (1-\hat{z})_+}
\nonumber\\
&&-\delta(1-\hat{z})(\ln{\hat{x}\over 1-\hat{x}}+3)\Bigr)
-{1\over 2\hat{z}}(1-2\hat{x})
\Bigr]\Biggr\}
\Biggr],
\label{realcross}
\eeq
where we performed partial integral,
\beq
\int^{1}_{x_B}dx{d\over dx}G_F(x,x)(1+\hat{x}^2)=\int^1_{x_B}{dx\over x}G_F(x,x)
(2\hat{x}^2-2\delta(1-\hat{x})),
\eeq
and we used $G_F(x,x_B)\delta(1-\hat{x})=G_F(x,x)\delta(1-\hat{x})$. 
The boundary condition of the integrals is determined by the condition 
$0<P_{h\perp}=\sqrt{{z^2(1-\hat{x})(1-\hat{z})\hat{z}\over \hat{x}}Q^2}< P^{\rm max}_{h\perp}$.
The hard cross sections (\ref{hard3})-(\ref{hard4}) associated with 
$\tilde{G}(x,x_B)$, $G(x_B,x_B-x)$ and $\tilde{G}(x_B,x_B-x)$ are new results derived 
in this study and, in particular, the latter two contributions came from the HP2 contribution~\cite{KT2009}
which was not
discussed in previous studies of the $P_{h\perp}$-weighted SSA. 
We should not neglect these new contributions
to demonstrate the cancellation of the collinear singularities. 
Other contributions (\ref{hard1})-(\ref{hard2}) agree with 
those derived in the previous study~\cite{KVX2012}.


\section{LO cross section and virtual-correction contribution in NLO cross section}

In this section, we introduce the results of the LO cross section and 
virtual-correction contribution in NLO cross section already derived in~\cite{KVX2012}.
Both contributions can be represented with $2\to 1$ scattering cross section.
\begin{figure}[h]
\begin{center}
  \includegraphics[height=4cm,width=10cm]{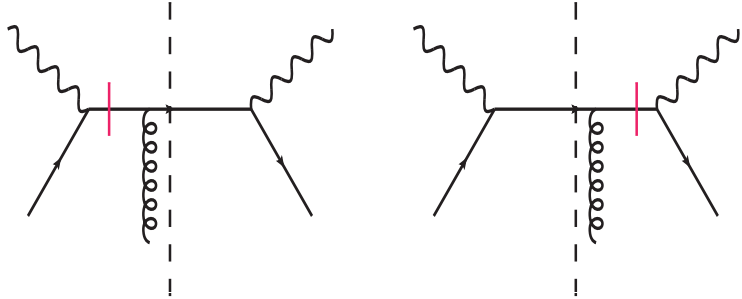}\hspace{1cm}
\end{center}
 \caption{Leading order diagram for the $P_{h\perp}$-weighted cross section.
 Left(Right) figure represents $S_{L(R)\gamma}(k_1,k_2)$.}
\end{figure}
The phase-space integral should be changed from $2\to 2$ scattering as follows.
\beq
&&{d^{3}p_c\over (2\pi)^{3}2p^{0}_c}
{d^{3}p_d\over (2\pi)^{3}2p^{0}_d}
(2\pi)^4\delta^4(xp+q-p_c-p_d)
\nonumber\\
&=&{d^{3}p_c\over (2\pi)^{3}2p^{0}_c}(2\pi)\delta\Bigl((xp+q-p_c)^2\Bigr)
\nonumber\\
&\to&{d^{3}p_c\over (2\pi)^{3}2p^{0}_c}
(2\pi)^4\delta^4(xp+q-p_c)
\eeq
In this case, we perform $P_{h\perp}$-integration before the collinear expansion as
\beq
&&\int d^{2}P_{h\perp}\epsilon^{\alpha\beta pn}S_{\perp\alpha}P_{h\perp\beta}
\Bigl(S_{L\gamma}(k_1,k_2)\delta^{2}(k_{2\perp}-{P_{h\perp}\over z})
+S_{R\gamma}(k_1,k_2)\delta^{2}(k_{1\perp}-{P_{h\perp}\over z})
\Bigr)
\nonumber\\
&=&\epsilon^{\alpha\beta pn}S_{\perp\alpha}z^3
\Bigl(k_{2\perp\beta}S_{L\gamma}(k_1,k_2)
+k_{1\perp\beta}S_{R\gamma}(k_1,k_2)
\Bigr),
\eeq
where we used the fact the virtual photon doesn't have transverse momentum in hadron frame.
We can find the following relation for LO diagram shown in Fig.6.
\beq
S_{Lp}(x_1p,x_2p)=-S_{Rp}(x_1p,x_2p)\equiv S_p(x_1p,x_2p).
\label{wardLO}
\eeq 
Since the $P_{h\perp}$-integration brought $O(k_{\perp})$ term,
the leading term of the collinear expansion gives twist-3 contribution.
We can construct the gluon's field strength tensor as follows.
\beq
&&\int{d^4k_1\over (2\pi)^4}\int{d^4k_2\over (2\pi)^4}
e^{ik_1\cdot \xi}e^{i(k_2-k_1)\cdot \eta}A^n(\eta)(k_{2\perp\beta}-k_{1\perp\beta})
S_{p}((k_1\cdot n)p, (k_2\cdot n)p) 
\nonumber\\
&=&i\int{d^4k_1\over (2\pi)^4}\int{d^4k_2\over (2\pi)^4}
e^{ik_1\cdot \xi}e^{i(k_2-k_1)\cdot \eta}F_{\beta}^{\ n}(\eta)S_{p}((k_1\cdot n)p, (k_2\cdot n)p)
\nonumber\\
&&+\int{d^4k_1\over (2\pi)^4}\int{d^4k_2\over (2\pi)^4}
e^{ik_1\cdot \xi}e^{i(k_2-k_1)\cdot \eta}A^{\perp}_{\beta}(\eta)
(k_2\cdot n-k_1\cdot n)S_{p}((k_1\cdot n)p, (k_2\cdot n)p).
\eeq
The last term vanishes due to the SGP delta function. Then we can use
the following formula for LO contribution
\beq
{d^4\la P_{h\perp}\Delta\sigma\ra^{\rm LO}\over dx_BdQ^2dz_hd\phi}
&=&{\alpha^2_{em}\over 8z_hx_B^2S^2_{ep}Q^2}
\int dzzD(z)\int dx_1\int dx_2
\epsilon^{S_{\perp}\alpha pn}iM_{Fij\,\alpha}(x_1,x_2)
\nonumber\\
&&\times\Bigl(-g_{\mu\nu}H_{ij\,p}^{\mu\nu}(x_1p,x_2p)\Bigr)
\Bigl({-2x\hat{x}\over \hat{u}Q^2}\Bigr)\delta(x_1-x_2)\delta(1-\hat{x})\delta(1-\hat{z}),
\eeq
which agrees with the corresponding formula in~\cite{VY2009,KVX2012}.
LO and NLO contributions in SIDIS were already calculated in previous work~\cite{KVX2012}. 
We just introduce their results in our notation below.
\beq
{d^4\la P_{h\perp}\Delta\sigma\ra^{\rm LO}\over dx_BdQ^2dz_hd\phi}
=-{z_h\pi M_N\alpha^2_{em}\over 4x_B^2S^2_{ep}Q^2}\sum_qe^2_qG^q(x_B,x_B)D^q(z_h)
\label{LOcross}
\eeq
\beq
{d^4\la P_{h\perp}\Delta\sigma\ra^{\rm virtual}\over dx_BdQ^2dz_hd\phi}
&=&-{z_h\pi M_N\alpha^2_{em}\over 4x_B^2S^2_{ep}Q^2}{\alpha_s\over 2\pi}\sum_qe_q^2
G^q(x_B,x_B)D^q(z_h)
\nonumber\\
&&\times\Bigl[C_F\Bigr({4\pi\mu^2\over Q^2}\Bigr)^{\epsilon}{1\over \Gamma(1-\epsilon)}
\Bigl(-{2\over \epsilon^2}-{3\over \epsilon}-8\Bigr)
\Bigr]
\label{virtualcross}
\eeq
Combining (\ref{realcross}), (\ref{LOcross}) and (\ref{virtualcross}), we obtain
the following complete formula for NLO $P_{h\perp}$-weighted cross section.
\beq
&&{d^4\la P_{h\perp}\Delta\sigma\ra^{\rm LO+NLO}\over dx_BdQ^2dz_hd\phi}
\nonumber\\
&=&-{\pi z_hM_N\alpha^2_{em}\over 4x_B^2S^2_{ep}Q^2}\sum_qe_q^2
\Biggl[G^q_F(x_B,x_B)D^q(z_h)
+{\alpha_s\over 2\pi}
\Bigl({4\pi\mu^2\over Q^2}\Bigr)^{\epsilon}{1\over \Gamma(1-\epsilon)}
\Bigl(-{1\over \epsilon}\Bigr)
\nonumber\\
&&\times\Biggl\{
D^q(z_h)\Bigl\{\int^1_{x_B}{dx\over x}\Bigl[P_{qq}(\hat{x})G^q_F(x,x)
+{N\over 2}\Bigl({(1+\hat{x})G^q_F(x_B,x)-(1+\hat{x}^2)G^q_F(x,x)\over (1-\hat{x})_+}
+\tilde{G}^q_F(x_B,x)\Bigr)\Bigr]
\nonumber\\
&&-NG^q_F(x_B,x_B)
+{1\over 2N}\int^1_{x_B}{dx\over x}\Bigl((1-2\hat{x})G^q_F(x_B,x_B-x)+\tilde{G}^q_F(x_B,x_B-x)\Bigr)
\Bigr\}
\nonumber\\
&&+G^q_F(x_B,x_B)
\int^1_{z_h}{dz\over z}P_{qq}(\hat{z})D^q(z)\Biggr\}
\nonumber\\
&&+{\alpha_s\over 2\pi}
\Bigl({4\pi\mu^2\over Q^2}\Bigr)^{\epsilon}{1\over \Gamma(1-\epsilon)}\int^1_{x_B}{dx\over x}
\int^1_{z_h}{dz\over z}
\Biggl\{x{dx\over x}G^q_F(x,x)D^q(z)
{1\over 2N\hat{z}}\Bigl[1-\hat{z}+{(1-\hat{x})^2+2\hat{x}\hat{z}\over (1-\hat{z})_+}
\nonumber\\
&&-\delta(1-\hat{z})\Bigl((1+\hat{x}^2)\ln{\hat{x}\over 1-\hat{x}}+2\hat{x}\Bigr)\Bigr]
+G^q_F(x,x)D^q(z){1\over 2N\hat{z}}\Bigl[
-2\delta(1-\hat{x})\delta(1-\hat{z})
\nonumber\\
&&+{2\hat{x}^3-3\hat{x}^2-1\over (1-\hat{x})_+(1-\hat{z})_+}
+{1+\hat{z}\over (1-\hat{x})_+}-2(1-\hat{x})
+\delta(1-\hat{z})\Bigl(
-(1-\hat{x})(1+2\hat{x})\log{\hat{x}\over 1-\hat{x}}
\nonumber\\
&&-2\Bigl({\ln(1-\hat{x})\over 1-\hat{x}}\Bigr)_+
+{2\over (1-\hat{x})_+}
-2(1-\hat{x})
+2{\ln\hat{x}\over (1-\hat{x})_+}
\Bigr)
+\delta(1-\hat{x})\Bigl(
(1+\hat{z})\ln\hat{z}(1-\hat{z})
\nonumber\\
&&-2{\ln\hat{z}\over (1-\hat{z})_+}
-2\Bigl({\ln(1-\hat{z})\over 1-\hat{z}}\Bigr)_+
+{2\hat{z}\over (1-\hat{z})_+}\Bigr)
\Bigr]
+G^q_F(x,x_B)D^q(z)\Bigl(C_F+{1\over 2N\hat{z}}\Bigr)\Bigl[
2\delta(1-\hat{x})\delta(1-\hat{z})
\nonumber\\
&&+{1+\hat{x}\hat{z}^2\over (1-\hat{x})_+(1-\hat{z})_+}
+\delta(1-\hat{z})\Bigl(
\log{\hat{x}\over 1-\hat{x}}+2\Bigl({\ln(1-\hat{x})\over 1-\hat{x}}\Bigr)_+
-2{\ln\hat{x}\over (1-\hat{x})_+}-{1+\hat{x}\over (1-\hat{x})_+}
\Bigr)
\nonumber\\
&&+\delta(1-\hat{x})\Bigl(
-(1+\hat{z})\ln\hat{z}(1-\hat{z})+2\Bigl({\ln(1-\hat{z})\over 1-\hat{z}}\Bigr)_+
+2{\ln\hat{z}\over (1-\hat{z})_+}-{2\hat{z}\over (1-\hat{z})_+}
\Bigr)
\Bigr]
\nonumber\\
&&+\tilde{G}^q_F(x,x_B)D^q(z)\Bigl(C_F+{1\over 2N\hat{z}}\Bigr)\Bigl[
-{1-\hat{x}\hat{z}^2\over (1-\hat{x})_+(1-\hat{z})_+}
+\delta(1-\hat{z})\Bigl(
\ln{\hat{x}\over 1-\hat{x}}+3
\Bigr)
\Bigr]
\nonumber\\
&&+G^q_F(x_B,x_B-x)D^q(z)\Bigl[{1\over 2N\hat{z}}\Bigl(
{(1-2\hat{x})\hat{z}^2\over (1-\hat{z})_+}
-\delta(1-\hat{z})(1-2\hat{x})
(\ln{\hat{x}\over 1-\hat{x}}+1)\Bigr)
\nonumber\\
&&+{1\over 2\hat{z}}(1-2\hat{x})\{(1-\hat{z})^2+\hat{z}^2\}
\Bigr]+\tilde{G}^q_F(x_B,x_B-x)D^q(z)\Bigl[{1\over 2N\hat{z}}\Bigl(
{\hat{z}^2\over (1-\hat{z})_+}
\nonumber\\
&&-\delta(1-\hat{z})(\ln{\hat{x}\over 1-\hat{x}}+3)\Bigr)
-{1\over 2\hat{z}}(1-2\hat{x})
\Bigr]-8C_F\delta(1-\hat{x})\delta(1-\hat{z})\Biggr\}
\Biggr],
\eeq
where $P_{qq}(x)$ is the splitting function
\beq
P_{qq}(x)=C_F\Bigl[{1+x^2\over (1-x)_+}+{3\over 2}\delta(1-x)\Bigr].
\eeq
The double pole terms ${2\over \epsilon^2}\delta(1-\hat{x})\delta(1-\hat{z})$ are cancelled
between the real cross section and the virtual cross section. The single pole term
in virtual cross section $2\times {3\over 2}\delta(1-\hat{x})\delta(1-\hat{z})$ is 
incorporated into the splitting functions.
The collinear singularities associated with the twist-3 functions can be subtracted 
with the following renormalization.
\beq
&&G_F(x_B,x_B)
\nonumber\\
&=&G^{(0)}_F(x_B,x_B)+{\alpha_s\over 2\pi}
\Bigr(-{1\over \hat{\epsilon}}\Bigr)\Bigl\{
\int^1_{x_B}{dx\over x}\Bigl[P_{qq}(\hat{x})G_F(x,x)
\nonumber\\
&&+{N\over 2}\Bigl({(1+\hat{x})G_F(x_B,x)-(1+\hat{x}^2)G_F(x,x)\over (1-\hat{x})_+}
+\tilde{G}_F(x_B,x)\Bigr)\Bigr]
-NG_F(x_B,x_B)
\nonumber\\
&&+{1\over 2N}\int^1_{x_B}{dx\over x}\Bigl((1-2\hat{x})G_F(x_B,x_B-x)+\tilde{G}_F(x_B,x_B-x)\Bigr)
\Bigr\},
\eeq
where we adopted the $\overline{{\rm MS}}$-scheme
\beq
{1\over \hat{\epsilon}}={1\over \epsilon}-\gamma_E+\ln4\pi.
\eeq
These collinear singularities are the same as those in F-type correlator (\ref{F-type}) 
at 1-loop order~\cite{BMP2009,MW2012,KQ2012}.
Then the collinear singularities are consistently subtracted and we can obtain the infrared-safe 
NLO cross section as follows.
\beq
&&{d^4\la P_{h\perp}\Delta\sigma\ra^{\rm LO+NLO}\over dx_BdQ^2dz_hd\phi}
\nonumber\\
&=&-{\pi z_hM_N\alpha^2_{em}\over 4x_B^2S^2_{ep}Q^2}\sum_qe_q^2
\Biggl[G^q_F(x_B,x_B,\mu)D^q(z_h,\mu)
\nonumber\\
&&+{\alpha_s\over 2\pi}\ln\Bigl({Q^2\over \mu^2}\Bigr)\Biggl\{D^q(z_h,\mu)\Bigl\{
\int^1_{x_B}{dx\over x}\Bigl[P_{qq}(\hat{x})G^q_F(x,x,\mu)
\nonumber\\
&&+{N\over 2}\Bigl({(1+\hat{x})G^q_F(x_B,x,\mu)-(1+\hat{x}^2)G^q_F(x,x,\mu)\over (1-\hat{x})_+}
+\tilde{G}^q_F(x_B,x,\mu)\Bigr)\Bigr]
\nonumber\\
&&-NG^q_F(x_B,x_B,\mu)
+{1\over 2N}\int^1_{x_B}{dx\over x}\Bigl((1-2\hat{x})G^q_F(x_B,x_B-x,\mu)
+\tilde{G}^q_F(x_B,x_B-x,\mu)\Bigr)
\Bigr\}
\nonumber\\
&&+G^q_F(x_B,x_B,\mu)
\int^1_{z_h}{dz\over z}P_{qq}(\hat{z})D^q(z,\mu)\Biggr\}
\nonumber\\
&&+{\alpha_s\over 2\pi}\int^1_{x_B}{dx\over x}\int^1_{z_h}{dz\over z}
\Biggl\{x{dx\over x}G^q_F(x,x,\mu)D^q(z,\mu)
{1\over 2N\hat{z}}\Bigl[1-\hat{z}+{(1-\hat{x})^2+2\hat{x}\hat{z}\over (1-\hat{z})_+}
\nonumber\\
&&-\delta(1-\hat{z})\Bigl((1+\hat{x}^2)\ln{\hat{x}\over 1-\hat{x}}+2\hat{x}\Bigr)\Bigr]
+G^q_F(x,x,\mu)D^q(z,\mu){1\over 2N\hat{z}}\Bigl[
-2\delta(1-\hat{x})\delta(1-\hat{z})
\nonumber\\
&&+{2\hat{x}^3-3\hat{x}^2-1\over (1-\hat{x})_+(1-\hat{z})_+}
+{1+\hat{z}\over (1-\hat{x})_+}-2(1-\hat{x})
+\delta(1-\hat{z})\Bigl(
-(1-\hat{x})(1+2\hat{x})\log{\hat{x}\over 1-\hat{x}}
\nonumber\\
&&-2\Bigl({\ln(1-\hat{x})\over 1-\hat{x}}\Bigr)_+
+{2\over (1-\hat{x})_+}
-2(1-\hat{x})
+2{\ln\hat{x}\over (1-\hat{x})_+}
\Bigr)
+\delta(1-\hat{x})\Bigl(
(1+\hat{z})\ln\hat{z}(1-\hat{z})
\nonumber\\
&&-2{\ln\hat{z}\over (1-\hat{z})_+}
-2\Bigl({\ln(1-\hat{z})\over 1-\hat{z}}\Bigr)_+
+{2\hat{z}\over (1-\hat{z})_+}\Bigr)
\Bigr]
\nonumber\\
&&+G^q_F(x,x_B,\mu)D^q(z,\mu)\Bigl(C_F+{1\over 2N\hat{z}}\Bigr)\Bigl[
2\delta(1-\hat{x})\delta(1-\hat{z})
+{1+\hat{x}\hat{z}^2\over (1-\hat{x})_+(1-\hat{z})_+}
\nonumber\\
&&+\delta(1-\hat{z})\Bigl(
\log{\hat{x}\over 1-\hat{x}}+2\Bigl({\ln(1-\hat{x})\over 1-\hat{x}}\Bigr)_+
-2{\ln\hat{x}\over (1-\hat{x})_+}-{1+\hat{x}\over (1-\hat{x})_+}
\Bigr)
\nonumber\\
&&+\delta(1-\hat{x})\Bigl(
-(1+\hat{z})\ln\hat{z}(1-\hat{z})+2\Bigl({\ln(1-\hat{z})\over 1-\hat{z}}\Bigr)_+
+2{\ln\hat{z}\over (1-\hat{z})_+}-{2\hat{z}\over (1-\hat{z})_+}
\Bigr)
\Bigr]
\nonumber\\
&&+\tilde{G}^q_F(x,x_B,\mu)D^q(z,\mu)\Bigl(C_F+{1\over 2N\hat{z}}\Bigr)\Bigl[
-{1-\hat{x}\hat{z}^2\over (1-\hat{x})_+(1-\hat{z})_+}
+\delta(1-\hat{z})\Bigl(
\ln{\hat{x}\over 1-\hat{x}}+3
\Bigr)
\Bigr]
\nonumber\\
&&+G^q_F(x_B,x_B-x,\mu)D^q(z,\mu)\Bigl[{1\over 2N\hat{z}}\Bigl(
{(1-2\hat{x})\hat{z}^2\over (1-\hat{z})_+}
-\delta(1-\hat{z})(1-2\hat{x})
(\ln{\hat{x}\over 1-\hat{x}}+1)\Bigr)
\nonumber\\
&&+{1\over 2\hat{z}}(1-2\hat{x})\{(1-\hat{z})^2+\hat{z}^2\}
\Bigr]+\tilde{G}^q_F(x_B,x_B-x,\mu)D^q(z,\mu)\Bigl[{1\over 2N\hat{z}}\Bigl(
{\hat{z}^2\over (1-\hat{z})_+}
\nonumber\\
&&-\delta(1-\hat{z})(\ln{\hat{x}\over 1-\hat{x}}+3)\Bigr)
-{1\over 2\hat{z}}(1-2\hat{x})
\Bigr]-8C_F\delta(1-\hat{x})\delta(1-\hat{z})\Biggr\}
\Biggr]+O(\alpha^2_s),
\eeq
where the scale dependence of $G_F(x,x,\mu^2)$ was introduced so that the cross section
doesn't depend on the artificial scale $\mu$. Then we can derive the scale evolution 
equation of $G_F(x,x,\mu^2)$ as
\beq
&&{\partial\over \partial\ln \mu^2}
{d^4\la P_{h\perp}\Delta\sigma\ra^{\rm LO+NLO}\over dx_BdQ^2dz_hd\phi}=0
\nonumber\\
&&\to
{\partial\over \partial\ln \mu^2}G_F(x_B,x_B,\mu^2)
={\alpha_s\over 2\pi}\Bigl\{
\int^1_{x_B}{dx\over x}\Bigl[P_{qq}(\hat{x})G_F(x,x,\mu^2)
\nonumber\\
&&+{N\over 2}\Bigl({(1+\hat{x})G_F(x_B,x,\mu^2)-(1+\hat{x}^2)G_F(x,x,\mu^2)\over (1-\hat{x})_+}
+\tilde{G}_F(x_B,x,\mu^2)\Bigr)\Bigr]
-NG_F(x_B,x_B,\mu^2)
\nonumber\\
&&+{1\over 2N}\int^1_{x_B}{dx\over x}\Bigl((1-2\hat{x})G_F(x_B,x_B-x,\mu^2)
+\tilde{G}_F(x_B,x_B-x,\mu^2)\Bigr)
\Bigr\}+O(\alpha^2_s),
\eeq
which completely agrees with the results in \cite{BMP2009,MW2012,KQ2012}.


\section{Summary}

We added the new hard pole contribution to the $P_{h\perp}$-weighted single-spin asymmetry
in semi-inclusive deep inelastic scattering.
Since the new pole contribution brings some collinear singularities at one-loop order, 
we should not neglect it for the exact cancellation of the collinear singularities.
Our result showed that the NLO $P_{h\perp}$-weighted cross section has
the same collinear singularities with the F-type correlator at one-loop order 
and then the singularities 
can be subtracted consistently. In addition, our calculation provided 
the scale evolution equation of the Qiu-Sterman function which completely agrees with
the corresponding results in different approaches.


\section*{Acknowledgments}

First the author would like to thank Zhong-Bo Kang for bringing his attention to their 
recent work~\cite{KVX2012}.
He also thanks Yuji Koike, Yoshitaka Hatta and Bo-Wen Xiao for 
helpful discussions and carefully reading his manuscript.
This work is supported in part by the NSFC under Grant No.~11575070.

\end{document}